\documentclass[twocolumn,aps,prb,epsf,showpacs,superscriptaddress]{revtex4}
\usepackage{amsmath}
\usepackage{amsfonts}
\usepackage{epsfig}
\usepackage{epsf}
\usepackage{array}

\begin{document}

\title{Global phase diagram of a doped Kitaev-Heisenberg model}

\author{Satoshi Okamoto}
\altaffiliation{okapon@ornl.gov}
\affiliation{Materials Science and Technology Division, Oak Ridge National Laboratory, Oak Ridge, Tennessee 37831, USA}

\begin{abstract}
The global phase diagram of a doped Kitaev-Heisenberg model is studied using an $SU(2)$ slave-boson mean-field method. 
Near the Kitaev limit, $p$-wave superconducting states which break the time-reversal symmetry are stabilized 
as reported by You {\it et al.} [Phys. Rev. B {\bf 86}, 085145 (2012)] 
irrespective of the sign of the Kitaev interaction. 
By further doping, a $d$-wave superconducting state appears when the Kitaev interaction is antiferromagnetic, 
while 
another $p$-wave superconducting state appears when the Kitaev interaction is ferromagnetic. 
This $p$-wave superconducting state does not break the time-reversal symmetry as reported by 
Hyart {\it et al.} [Phys. Rev. B {\bf 85}, 140510 (2012)], and such a superconducting state 
also appears when the antiferromagnetic Kitaev interaction and the ferromagnetic Heisenberg interaction compete. 
This work, thus, 
demonstrates the clear difference between the antiferromagnetic Kitaev model and the ferromagnetic Kitaev model
when carriers are doped while these models are equivalent in the undoped limit, 
and how novel superconducting states emerge when the Kitaev interaction and the Heisenberg interaction compete.  
\end{abstract}

\maketitle



\section{Introduction}

There has been considerable attention paid to the Kitaev model whose ground state is a gapless $Z_2$ 
spin liquid (SL).\cite{Kitaev06} 
If such a model is realized, 
fault tolerant quantum computations can be possible. 

\begin{figure}[tbp]
\includegraphics[width=0.8\columnwidth]{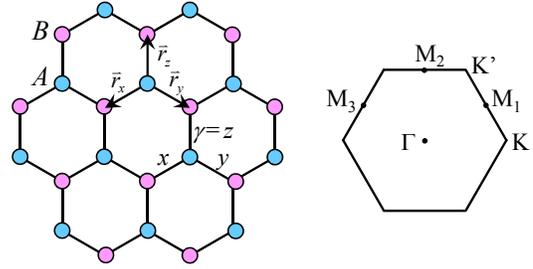}
\caption{(Color online) Schematic view of the Kitaev-Heisenberg model. 
$\gamma=x,y,z$ in the left figure show the spin components for the Kitaev interaction.
$\vec r_{x,y,z}$ are unit vectors connecting the nearest-neighbor sites. 
On the right, the first Brillouin zone is shown.}
\label{fig:structure}
\end{figure}

The Kitaev model consists of local (iso)spins $S=1/2$ on a honeycomb lattice as 
\begin{equation}
H_K = J_K \sum_{\langle \vec r \vec r' \rangle } S_{\vec r}^\gamma S_{\vec r'}^\gamma . 
\label{eq:HK}
\end{equation}
Here, the spin component $\gamma$ depends on the bond specie as shown in Fig.~\ref{fig:structure}. 
$A_2$IrO$_3$ ($A$=Li or Na) 
have been proposed as possible candidates to realize the Kitaev model 
as Ir$^{4+}$ ions having the effective angular momentum $j_{eff}=1/2$ 
form the honeycomb lattice.\cite{Chaloupka10} 
In fact, if the correlation effects are strong enough to realize a Mott insulating state, 
the low-energy electronic state is described by the combination of 
the anisotropic Kitaev interaction [Eq.~(\ref{eq:HK})] and the symmetric Heisenberg interaction, 
$H_J = J_H \sum_{\langle \vec r \vec r' \rangle } \vec S_{\vec r} \cdot \vec S_{\vec r'}$, 
called the Kitaev-Heisenberg (KH) model. 
Alternatively, 
density-functional-theory calculations for Na$_2$IrO$_3$
predicted the quantum spin Hall effect.\cite{Shitade09} 
Later experimental measurements for Na$_2$IrO$_3$ 
confirmed a magnetic long-range order with a ``zigzag'' antiferromagnetic (AFM) pattern.\cite{Singh10,Liu11} 
As this magnetic pattern is not realized in the model first proposed for Na$_2$IrO$_3$, 
where the Kitaev interaction was introduced as a ferromagnetic (FM) interaction ($J_K<0$) and 
the Heisenberg interaction was introduced as an AFM interaction ($J_H>0$),\cite{FMKitaevAFHeisenberg}
the importance of additional contributions such as longer-range magnetic couplings\cite{Kimchi11,Singh12,Choi12,Ye12} 
and lattice distortions\cite{Bhattacharjee12} were suggested. 
Recently, the sign of Kitaev and Heisenberg terms was reconsidered\cite{Chaloupka12} 
by including the direct hybridization between neighboring Ir $t_{2g}$ and $e_g$ orbitals.\cite{Khaliullin05} 
It is found that, when the Kitaev interaction is AFM and the Heisenberg interaction is FM, 
zigzag-type AFM ordering could be stabilized in accordance with the experimental report.

While the Kitaev SL state is not realized in Na$_2$IrO$_3$, 
there could appear novel states by carrier doping if this system is described by the KH model. 
Specifically, considering the FM Kitaev-AFM Heisenberg model, 
such an effect was studied in Refs.~\onlinecite{You11} and \onlinecite{Hyart12}. 
Both studies found the triplet ($p$) superconductivity (SC) by carrier doping, but 
the $SU(2)$ slave-boson mean-field (SBMF) study found a state which breaks the time-reversal symmetry (termed $p$ SC$_1$),\cite{You11}
while the $U(1)$ SBMF study found a time-reversal symmetric state (termed $p$ SC$_2$).\cite{Hyart12} 
Exotic triplet pairing was also suggested from a low-energy effective model for layered cobaltate.\cite{Khaliullin04} 
Recently, artificial bilayers of perovskite transition-metal oxides (TMOs) grown along the [111] crystallographic axis 
were proposed as new platforms to explore a variety of quantum effects.\cite{Xiao11} 
It was pointed out\cite{Okamoto12} that such a bilayer involving SrIrO$_3$ (Ref.~\onlinecite{Cao07}) 
could also realize the KH model when the correlation effects are strong enough to yield a Mott insulating state. 
But, both the Kitaev interaction and the Heisenberg interaction were found to be AFM. 
Doping carriers into such an AFM Kitaev-FM Heisenberg model was also shown to stabilize the $p$ SC$_1$ state but such a state 
becomes unstable against a singlet SC state by further doping. 
Doping effects in the general KH model have not been studied, 
including the AFM Kitaev-FM Heisenberg interaction as alternatively suggested for Na$_2$IrO$_3$.

In this paper, we consider a general KH model in which both Kitaev interaction and Heisenberg interaction can be 
either FM or AFM. 
Doping effects are considered by introducing hopping terms which conserve isospin index $\sigma$ with the double occupancy prohibited 
as in the $tJ$ model for high-$T_c$ cuprates. 
The Hamiltonian is thus given by 
\begin{eqnarray}
H \!\!&=&\!\! -t \sum_{\langle \vec r \vec r' \rangle } \bigl(c_{\vec r \sigma}^\dag c_{\vec r' \sigma} + H.c. \bigr) 
+ H_K + H_H. 
\label{eq:model}
\end{eqnarray}
We investigate the global phase diagram of this model using an $SU(2)$ SBMF method. 
We start from solving the undoped KH model defined on a finite cluster 
using the Lanczos exact diagonalization method. 
We then introduce a mean-field decoupling scheme that can be applied for both symmetric Heisenberg interaction and 
the anisotropic Kitaev interaction. 
Mean-field ans{\"a}tze are constructed motivated by such exact solutions. 
Our results demonstrate the clear difference between the AFM Kitaev model and the FM Kitaev model 
when carriers are doped, even though the undoped cases are equivalent. 
We confirmed novel triplet superconducting states reported previously. 
Yet, their relative stability is found to depend on the sign of the Kitaev interaction and 
the competition between the Kitaev interaction and the Heisenberg interaction. 
Additionally, $s$-wave and $d$-wave superconducting states in the AFM Heisenberg limit and 
the FM state in the FM Heisenberg limit are found. 
Our results could become guidelines for a materials search to realize specific properties 
and further theoretical analyses. 
As the present model is simple, 
testing or refining the current results by using more sophisticated methods is also possible and desirable.

The rest of this paper is organized as follows: 
In Sec.~\ref{sec:undoped}, we examine the undoped KH model 
by using the Lanczos exact diagonalization method. 
The results are useful for selecting mean-field ans{\"a}tze to be used later. 
A mean-field method is introduced in Sec.~\ref{sec:SBMF}, and 
our results are presented in Sec.~\ref{sec:results}. 
Section \ref{sec:summarydiscussion} is devoted to summary and discussion.

\section{Undoped case}
\label{sec:undoped}

\begin{figure*}[tbp]
\includegraphics[width=1.5\columnwidth]{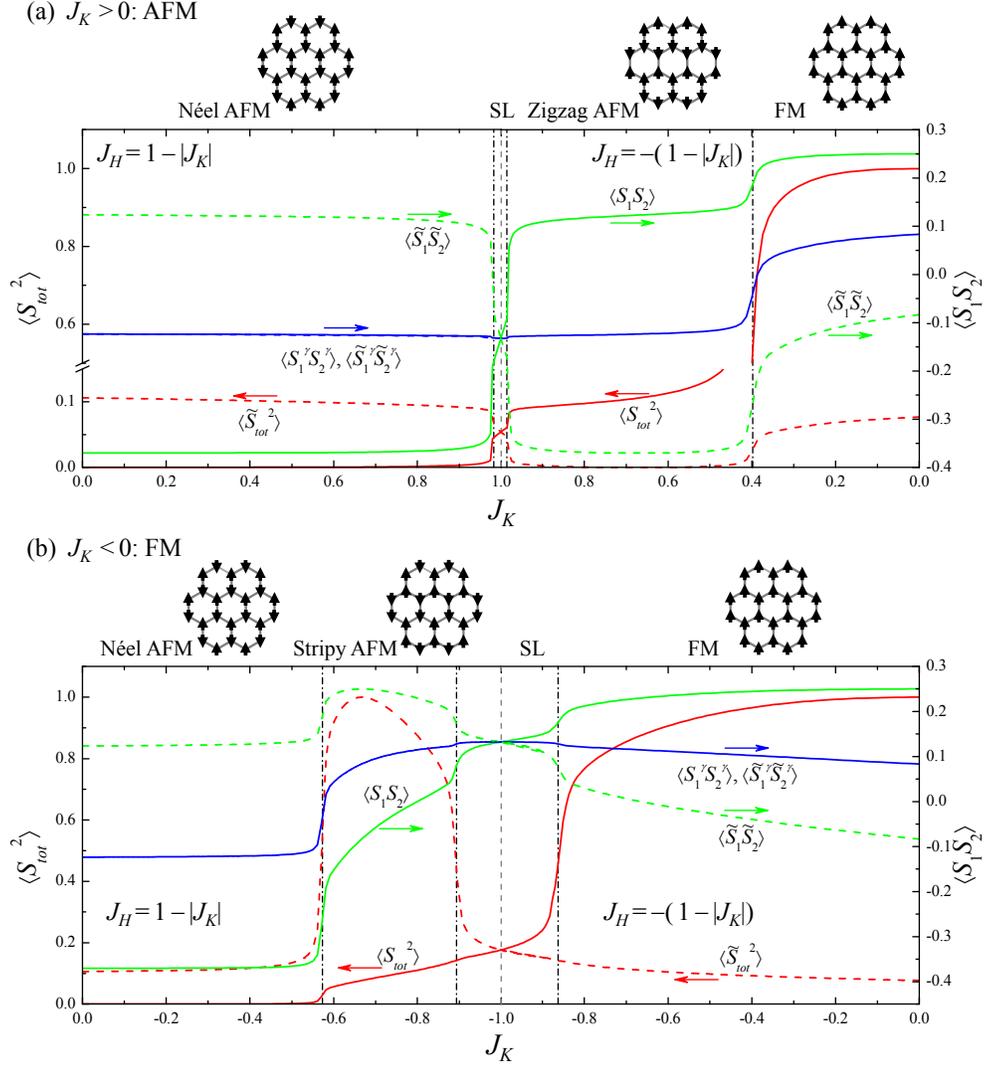}
\caption{(Color online) Lanczos exact diagonalization results for squared total spins 
(normalized to their values in the fully polarized FM state) 
and the NN spin correlations obtained on 24-site clusters as a function of $J_K$ 
with $J_H = \pm (1-|J_K|)$. 
(a) AFM Kitaev case $J_K>0$  and (b) FM Kitaev $J_K<0$. 
Solid (dashed) lines correspond to original (rotated) spin basis.
Vertical dash-dotted lines are first-order phase boundaries. 
For each case, a magnetic pattern is schematically shown. 
}
\label{fig:lanczos}
\end{figure*}

We first analyze the undoped KH model in detail. 
This analysis will be helpful for considering mean-field ans{\"a}tze and understanding the phases 
arising by carrier doping. 

Before going into the detailed analysis, 
it is instructive to perform the four-sublattice transformation.\cite{Chaloupka10,Khaliullin05} 
The four-sublattice transformation leads to the change in the sign of the Heisenberg term with 
$J_H \rightarrow - J_H$ and $J_K \rightarrow J_K + 2 J_H$. 
When the Kitaev term and the Heisenberg term have the different sign, $J_K$ vanishes at $J_K = -2 J_H$. 
As the resulting Heisenberg model is AFM for the FM Heisenberg case and FM for the AFM Heisenberg case, 
the spin ordering around 
$J_K= -2 J_H$ is ``zigzag AFM'' for the former and ``stripy AFM'' for the latter. 
Due to the larger quantum fluctuation, the total ``staggered spin'' in the rotated spin basis is reduced, 
and the parameter regime for this zigzag AFM is expected to be wider than that for the stripy AFM. 
When the Kitaev term and the Heisenberg term have the same sign, 
the cancellation does not occur in $J_K$. 
Thus, a direct transition is expected between 
the Kitaev SL in the large $|J_K|$ regime and other competing phase stabilized in the large $|J_H|$ regime: 
the N{\'e}el AFM or the FM.

We now employ the Lanczos exact diagonalization for the Hamiltonian for the undoped KH model $H_K+H_H$ 
defined on a 24-site cluster with periodic boundary condition. 
This cluster is compatible with the four-sublattice transformation\cite{Chaloupka10} 
which changes the original spin $S$ to $\widetilde S$. 
Numerical results for squared total spin and the nearest-neighbor (NN) spin correlations are shown in Fig.~\ref{fig:lanczos}. 
As expected, there are two phases, N{\'e}el AFM and SL, for the AFM Kitaev-AFM Heisenberg case (upper left) and 
three phases, FM, zigzag AFM, and SL, for the AFM Kitaev-FM Heisenberg case (upper right). 
The phase boundaries are also signaled as peaks in the second derivatives of the total energy (not shown). 
In both cases, 
the SL regime is rather narrow with the nearly identical critical value $|J_{K,c}| \sim 0.98$ separating 
it from magnetically ordered phases. 
For the AFM Kitaev-FM Heisenberg case, the phase boundary between zigzag AFM and FM is shifted 
from the classical value $J_K=1/2$ to a smaller value $J_K \sim 0.4$ as discussed above. 
For the FM Kitaev case, 
the situation is just opposite to the AFM Kitaev case 
with the N{\'e}el ordering replaced by FM and the zigzag AFM by the stripy AFM.
Here, the phase boundary between the N{\'e}el AFM and the stripy AFM is shifted from the classical value $J_K= - 1/2$ 
to $J_K \sim - 0.57$. 

It is noted that the AFM Kitaev interaction is more destructive for the FM ordering than 
the FM Kitaev interaction for the FM ordering. 
All phase boundaries are consistent with the recent report in Ref.~\onlinecite{Chaloupka12}
as obtained from the second derivative of the total energy.

\section{Slave-boson mean-field theory}
\label{sec:SBMF}

In this section, we introduce a SBMF method that can be applied for both Heisenberg and Kitaev interactions. 
As usual, an $S=1/2$ isospin operator is described by fermionic spinons $f_{\sigma}$ as 
$S_{\vec r}^\gamma = \frac{1}{2} f_{\vec r \sigma}^\dag \tau_{\sigma \sigma'}^\gamma f_{\vec r \sigma'}$ 
with the local constraint $\sum_\sigma f_{\vec r \sigma}^\dag f_{\vec r \sigma} =1$, 
which is normally approximated as the global constraint. 
$\hat \tau^\gamma$ is a Pauli matrix. 

In order to deal with the doping effect near a Mott insulating state excluding the double occupancy, 
two bosonic auxiliary particles $b_{1,2}$ are introduced as 
$c_{\vec r \sigma} \Rightarrow \frac{1}{\sqrt{2}}
(b^\dag_{\vec r 1} f_{\vec r \sigma} + \sigma b^\dag_{\vec r 2} f_{\vec r \bar \sigma}^\dag )$ (Ref.~\onlinecite{Lee06}) 
with the $SU(2)$ singlet condition\cite{You11}
\begin{eqnarray}
K_{\vec r}^\gamma = \frac{1}{4} {\rm Tr} \, 
F_{\vec r} \hat \tau^\gamma F^\dag_{\vec r} 
- \frac{1}{4} {\rm Tr} \, \hat \tau^z B^\dag_{\vec r} \hat \tau^\gamma B_{\vec r}= 0, 
\end{eqnarray}
with 
\begin{eqnarray}
F_{\vec r} = 
\Biggl({f_{\vec r \uparrow} \atop f_{\vec r \downarrow}} 
{-f^\dag_{\vec r \downarrow} \atop f^\dag_{\vec r \uparrow}} \Biggr), 
B_{\vec r} = \Biggl({b^\dag_{\vec r 1} \atop b^\dag_{\vec r 2}} {-b_{\vec r 2} \atop b_{\vec r 1}} \Biggr) .
\end{eqnarray}
The global constraints $\langle K^\gamma \rangle=0$ are imposed by $SU(2)$ gauge potentials $a^\gamma$. 
Doped carriers can be either holes or electrons. 
As the current model has only NN hoppings [see Eq.~(\ref{eq:model})], 
there exists particle-hole symmetry about the zero doping, therefore the effect is symmetric. 
Focusing on the low-doping regime at zero temperature, 
we assume that all bosons are condensed, so that 
$\delta = \sum_\nu \langle b_{\nu \vec r}^\dag b_{\nu \vec r}\rangle 
\approx \sum_\nu |\langle b_{\nu \vec r} \rangle|^2 $ and 
$\langle b_{\nu \vec r \in A} \rangle = (\pm i) \langle b_{\nu \vec r' \in B} \rangle$. 
Imaginary number $i$ appears when the Bose condensation acquires the sublattice-dependent phase.\cite{You11}

\subsection{Decoupling scheme}

In order to apply the SBMF method for both AFM and FM Kitaev interactions and AFM and FM Heisenberg interactions, 
we employ the decoupling scheme introduced in Ref.~\onlinecite{Okamoto12}. 
Here, a spin quadratic term is decoupled into several different channels as 
\begin{eqnarray}
S_{\vec r}^\gamma S_{\vec r'}^\gamma \!\!\! &=& \!\!\!
-\frac{1}{8} \bigl(\Delta_{\vec r \vec r'}^* \Delta_{\vec r \vec r'} + \chi_{\vec r \vec r'}^*\chi_{\vec r \vec r'} 
+ t_{\vec r \vec r'}^{\gamma *}t_{\vec r \vec r'}^\gamma 
+ e_{\vec r \vec r'}^{\gamma *}e_{\vec r \vec r'}^\gamma \bigr) \nonumber \\
&& +\frac{1}{8} \sum_{\gamma' \ne \gamma} \bigl( t_{\vec r \vec r'}^{\gamma' *}t_{\vec r \vec r'}^{\gamma'} 
+ e_{\vec r \vec r'}^{\gamma' *}e_{\vec r \vec r'}^{\gamma'} \bigr), 
\label{eq:decoupling}
\end{eqnarray}
where 
$\Delta_{\vec r \vec r'} = f_{\vec r \sigma} i \tau^y_{\sigma \sigma'} f_{\vec r' \sigma'} $ (singlet pairing), 
$t^\gamma_{\vec r \vec r'} = f_{\vec r \sigma} [i \hat \tau^\gamma \hat \tau^y]_{\sigma \sigma'} f_{\vec r' \sigma'} $ 
(triplet pairing), 
$\chi_{\vec r \vec r'} = f_{\vec r \sigma}^\dag f_{\vec r' \sigma'}$ (spin-conserving exchange term), and 
$e^\gamma_{\vec r \vec r'} = f_{\vec r \sigma}^\dag \tau^\gamma_{\sigma \sigma'} f_{\vec r' \sigma'}$ (spin-nonconserving exchange term). 
Summation over $\gamma$ in Eq.~(\ref{eq:decoupling}) gives a Heisenberg term. 
Then, terms having the negative coefficient are kept and 
the mean field decoupling is introduced to them. 
This recovers the previous mean-field schemes.\cite{Lee06,Shindou09,Schaffer12}
Different decoupling schemes are also used in literature.\cite{Burnell11,You11,Hyart12}

In what follows, we use the simplified notation in which the subscript $\vec r \vec r'$ is replaced by the bond index 
$\rho=x, y, z$ connecting the sites $\vec r \in A$ and $\vec r' \in B$, 
for example, 
$\langle \chi_{\vec r \vec r'} \rangle$ for $\vec r'- \vec r=\vec r_\rho$ is written as $\langle \chi_\rho \rangle$. 
$\vec r_\rho$ is a unit vector connecting the nearest-neighboring sites along the $\rho$ bond as shown in Fig.~\ref{fig:structure}. 
These are explicitly given by 
$\vec r_x = (-\sqrt{3}/2,-1/2)$, $\vec r_y = (\sqrt{3}/2,-1/2)$ and 
$\vec r_z = (0,1)$.

\subsection{Mean field Hamiltonian}

After the mean-field decoupling, the single-particle Hamiltonian 
is expressed as 
\begin{equation}
H^{MF}=\sum_{\vec k} \sum_{\sigma \sigma'} \varphi_{\vec k \sigma}^\dag \bigl\{\hat H_t (\vec k) + \hat H_K (\vec k) + \hat H_H (\vec k) \bigr\} 
\varphi_{\vec k \sigma'}
+H_0.
\label{eq:HMF}
\end{equation}
Here, a Nambu representation is used with 
4-component spinors $\varphi^\dag_{\vec k \sigma}$ given by 
$\varphi^\dag_{\vec k \sigma} = 
\bigl(f^\dag_{\vec k A \sigma}, f^\dag_{\vec k B \sigma}, f_{-\vec k A \sigma}, f_{-\vec k B \sigma}\bigr)$. 
$\hat H_{t,K,H}$ are $8 \times 8$ matrices. 
$\hat H_{t}$ includes both hopping terms and the chemical potential or the gauge field and is given by 
\begin{widetext}
\begin{eqnarray}
\hat H_{t} (\vec k) = 
\left[
\begin{array}{cccc}
- a^z \, \delta_{\sigma \sigma'} & \chi (\vec k) \delta_{\sigma \sigma'} & (a^x + i a^y) \varepsilon_{\sigma \sigma'} 
&  \\
\chi^* (\vec k) \delta_{\sigma \sigma'} & - a^z \, \delta_{\sigma \sigma'} & & 
(a^x + i a^y) \varepsilon_{\sigma \sigma'} \\
(a^x - i a^y) \varepsilon_{\sigma' \sigma}&  & a^z \, \delta_{\sigma \sigma'} & 
-\chi^* (-\vec k) \delta_{\sigma \sigma'}\\
  & (a^x - i a^y) \varepsilon_{\sigma' \sigma} & -\chi^* (-\vec k) \delta_{\sigma \sigma'} & 
a^z \, \delta_{\sigma \sigma'}
\end{array}
\right], \label{eq:Ht}
\end{eqnarray}
where 
\begin{equation}
\chi (\vec k) = - \frac{1}{2} \sum_\rho e^{i \vec k \cdot \vec r_\rho} 
\delta (i) t ,
\end{equation}
and 
$\varepsilon_{\uparrow \downarrow}=-\varepsilon_{\downarrow \uparrow} = 1$ is the antisymmetric tensor. 
The prefactor $\frac{1}{2} \delta (i)$ for $t$ comes from the mean-field decoupling for the bosonic term 
$\langle b_{A 1} b^\dag_{B 1} -  b^\dag_{A 2} b_{B 2}  \rangle$. 

Spin-spin interaction terms are both expressed as 
\begin{eqnarray}
\hat H_{K,H} (\vec k) = 
|J_{K,H}| \left[
\begin{array}{cccc}
  & \chi_{\sigma \sigma'} (\vec k) &   
& \Delta_{\sigma \sigma'} (\vec k) \\
\chi^*_{\sigma' \sigma} (\vec k) &   & -\Delta_{\sigma' \sigma} (-\vec k) & 
  \\
 & -\Delta^*_{\sigma \sigma'} (-\vec k) &   & 
-\chi^*_{\sigma' \sigma} (-\vec k)\\
\Delta^*_{\sigma' \sigma} (\vec k) &   & -\chi^*_{\sigma \sigma'} (-\vec k) & 
\end{array}
\right]. 
\end{eqnarray}
\end{widetext}
For the AFM Kitaev interaction, the matrix elements are given by  
\begin{eqnarray}
\hat \chi (\vec k) \!\!&=&\!\! - \frac{1}{8} \sum_\rho e^{i \vec k \cdot \vec r_\rho} 
\langle \chi_\rho^* \rangle \hat \tau^0 
- \frac{1}{8} \sum_{\rho} e^{i \vec k \cdot \vec r_\rho} 
\langle e^{\rho *}_\rho \rangle
\hat \tau^\rho ,
\label{eq:chiAFKitaev}\\
\hat \Delta (\vec k) \!\!&=&\!\! \frac{1}{8} \sum_{\rho} e^{i \vec k \cdot \vec r_\rho}
\langle \Delta_\rho \rangle i \hat \tau^y 
- \frac{1}{8} \sum_{\rho} e^{i \vec k \cdot \vec r_\rho} 
\langle t^{\rho}_\rho \rangle 
 i \hat \tau^y \hat \tau^\rho , \quad
\label{eq:DeltaAFKitaev}
\end{eqnarray}
with $\hat \tau^0$ being the $2 \times 2$ unit matrix and, for the FM Kitaev interaction, these are given by 
\begin{eqnarray}
\hat \chi (\vec k) \!\!&=&\!\! 
- \frac{1}{8} \sum_{\rho \gamma} (1- \delta_{\rho \gamma}) e^{i \vec k \cdot \vec r_\rho} 
\langle e^{\gamma *}_\rho \rangle
\hat \tau^\gamma ,
\label{eq:chiFMKitaev}\\
\hat \Delta (\vec k) \!\!&=&\!\! 
- \frac{1}{8} \sum_{\rho \gamma} (1- \delta_{\rho \gamma}) e^{i \vec k \cdot \vec r_\rho} 
\langle t^{\gamma}_\rho \rangle 
i \hat \tau^y \hat \tau^\gamma.  
\label{eq:DeltaFMKitaev}
\end{eqnarray}
For the AFM Heisenberg interaction, we have the well known expressions 
\begin{eqnarray}
\hat \chi (\vec k) \!\!&=&\!\! - \frac{3}{8} \sum_\rho e^{i \vec k \cdot \vec r_\rho} 
\langle \chi_\rho^* \rangle \hat \tau^0 , 
\label{eq:chiAFHeiisenberg}\\
\hat \Delta (\vec k) \!\!&=&\!\! \frac{3}{8} \sum_{\rho} e^{i \vec k \cdot \vec r_\rho}
\langle \Delta_\rho \rangle i \hat \tau^y ,
\label{eq:DeltaAFHeiisenberg}
\end{eqnarray}
while, for the FM Heisenberg, we have 
\begin{eqnarray}
\hat \chi (\vec k) \!\!&=&\!\! 
- \frac{1}{8} \sum_{\rho \gamma} e^{i \vec k \cdot \vec r_\rho} 
\langle e^{\gamma *}_\rho \rangle 
\hat \tau^\gamma ,
\label{eq:chiFMHeiisenberg}\\
\hat \Delta (\vec k) \!\!&=&\!\! 
- \frac{1}{8} \sum_{\rho \gamma} e^{i \vec k \cdot \vec r_\rho} 
\langle t^{\gamma}_\rho \rangle 
i \hat \tau^y \hat \tau^\gamma.  
\label{eq:DeltaFMHeiisenberg}
\end{eqnarray}

$H_0$ is a constant term 
for which the contributions from the AFM Kitaev and the FM Kitaev are given by 
$\frac{1}{8}\sum_{\rho} |J_K| \bigl(|\langle \chi_\rho \rangle |^2 + |\langle e^\rho_\rho \rangle|^2 
+ |\langle \Delta_\rho \rangle |^2 + |\langle t^\rho_\rho \rangle |^2 \bigr)$ 
and 
$\frac{1}{8}\sum_{\rho \gamma} |J_K| (1-\delta_{\rho \gamma}) \bigl(|\langle e^\rho_\rho \rangle|^2 
+ |\langle t^\rho_\rho \rangle |^2 \bigr)$, 
respectively, 
and the contributions from the AFM Heisenberg and the FM Heisenberg are given by 
$\frac{3}{8} \sum_{\rho} |J_H| \bigl(|\langle \chi_\rho \rangle|^2 + |\langle \Delta_\rho \rangle|^2 \bigr)$
and
$\frac{1}{8} \sum_{\rho \gamma} |J_H| \bigl(|\langle e^\gamma_\rho \rangle|^2 + |\langle t^\gamma_\rho \rangle|^2 \bigr)$, 
respectively.

Mean-field Hamiltonians shown in this subsection might become useful for refining the results to be presented 
by using variational techniques. 
In principle, one can construct variational wave functions by 
(1) diagonalizing mean-field single-particle Hamiltonians without contributions from slave bosons 
and (2) projecting out the unphysical doubly occupied states. 
Then, the total energy is computed by using thus constructed variational wave functions 
and is minimized with respect to variational parameters.

\subsection{Mean-field ans{\"a}tze}

{\it Undoped Kitaev limit}.
The undoped FM Kitaev model was studied using the SBMF theory in Ref.~\onlinecite{Schaffer12}, 
and the undoped AFM Kitaev model was studied in Ref.~\onlinecite{Okamoto12}. 
As expected from the true ground state of the Kitaev model which does not depend on the signs of exchange constants,\cite{Kitaev06}
the two cases are shown to give the identical excitation spectrum. 

Using the current definition, 
the mean-field solution for the FM Kitaev model is given by 
$- i \langle t_{x}^y \rangle = - i \langle t_{z}^y \rangle 
= \langle t^x_y\rangle = \langle t^x_z\rangle = \langle e^z_x\rangle = \langle e^z_y\rangle = 0.3812 i$ and 
$\langle t^x_x\rangle = -i \langle t^y_y\rangle = \langle e^z_z\rangle = -0.1188 i$. 
The mean-field solution for the AFM Kitaev model is given by 
$- \langle \chi_{x,y,z} \rangle = - \langle e^z_z\rangle = \langle t^x_x\rangle = i \langle t^y_y\rangle = 0.3812 i$ and 
$- \langle e^z_x\rangle = - \langle e^z_y\rangle = \langle t^x_y\rangle = \langle t^x_z\rangle 
= i \langle t^y_x\rangle = i \langle t^y_z\rangle = -0.1188 i$ 

As discussed in detail in Ref.~\onlinecite{Schaffer12}, the first mean-field ansatz describes a $Z_2$ SL. 
The second ansatz uses the same gauge used in Refs.~\onlinecite{You11} and \onlinecite{Schaffer12}, 
where the dispersive Majorana fermion mode is given by $\chi^0 = \frac{1}{\sqrt{2}}(f_\uparrow + f_\uparrow^\dag)$. 
Thus, the mean-field ansatz for the AFM Kitaev model also describes a $Z_2$ SL.

In doped cases, a mean-field Hamiltonian has additional three gauge potentials. 
With possible magnetic orderings, a total of $\sim 30$ parameters have to be determined self-consistently. 
In order to make the problem tractable, 
we focus on the following five ans{\"a}tze.
The first four ans{\"a}tze respect the sixfold rotational symmetry of the underlying lattice. 

{\it $p$ SC$_1$}. 
This mean-field ansatz is adiabatically connected to the mean-field solution for the Kitaev limit as described above. 
Here, the relative phase $\pm i$ is required between the Bose condensation at sublattices $A$ and $B$ 
with the $SU(2)$ gauge potentials $a^x=a^y=a^z$.\cite{You11} 
Because of this constraint, 
the spinon density $\langle f_{\vec r \sigma}^\dag f_{\vec r \sigma} \rangle$ differs from 
the ``real'' electron density $\langle c_{\vec r \sigma}^\dag c_{\vec r \sigma}\rangle$ 
in the $p$ SC$_1$ phase and a normal phase 
($\langle t_\rho^\gamma \rangle = \langle e_\rho^\gamma \rangle = \langle \Delta_\rho \rangle =0$) 
adjacent to it. 
In many cases, such a normal phase has slightly lower energy than the other SC ans{\"a}tze, 
but this is an artifact of the constraint. 
In this work, we identify the upper bound for the $p$ SC$_1$ phase as the smaller $\delta$ where 
the order parameters for the $p$ SC$_1$ phase become zero or the $p$ SC$_1$ phase becomes higher in energy than the other phases.

{\it $p$ SC$_2$}.
The second ansatz is also a $p$ SC. 
We assume the form of order parameters based on the leading pairing instability in the stability matrices 
$M_{x,y,z}$ (Refs.~\onlinecite{Black07,Hyart12}) as 
$\mbox{\boldmath $d$} = 
\mbox{\boldmath $d$}_x + \mbox{\boldmath $d$}_y + \mbox{\boldmath $d$}_z$.
Here, 
$\mbox{\boldmath $d$}_\gamma = \langle t^\gamma_x, t^\gamma_y, t^\gamma_z \rangle$, 
%
and we take 
$\mbox{\boldmath $d$}_x = e^{i \theta_x} ( t_1, t_2, t_3 )$, 
$\mbox{\boldmath $d$}_y = e^{i \theta_y} ( t_3, t_1, t_2 )$, 
and $\mbox{\boldmath $d$}_z = e^{i \theta_z} ( t_2, t_3, t_1 )$ 
with $t_{1,2,3}$ being real. 
All solutions with $\theta_\gamma - \theta_{\gamma'}=0$ or $\pi$ for $\gamma \ne \gamma'$ are found to degenerate 
and are lower in energy than the other combinations for both the AFM Kitaev and the FM Kitaev cases 
as reported in Ref.~\onlinecite{Hyart12}. 
The details of the stability matrices and the symmetry of the order parameters are given 
in Appendix \ref{app:stabilitymatrix}.

{\it $s$ SC}. 
The third ansatz is a singlet SC with the $s$ wave paring. 
The SC order parameter is symmetric as $\langle \Delta_x \rangle= \langle \Delta_y \rangle = \langle \Delta_z \rangle = \Delta$. 

{\it $d+id$ SC}. 
The fourth ansatz is also a singlet SC with the $d_{x^2-y^2} + i d_{xy}$ pairing (in short $d+id$ pairing).\cite{Black07}  
The spatial dependence of the SC order parameter is given by 
$\langle \Delta_x, \Delta_y, \Delta_z \rangle = \Delta (e^{-2 \pi i/3}, e^{2 \pi i/3}, 1)$.

For the latter three ans{\"a}tze, we further introduce the following conditions: 
(1) Order parameters $\langle e^\gamma_\rho \rangle$ are assumed to be zero because 
these indeed become zero at large dopings and 
the fermionic dispersion relations generally break the hexagonal symmetry 
when both $\langle e^\gamma_\rho \rangle$ and pairing order parameters $\langle t^\gamma_\rho \rangle$ or 
$\langle \Delta_\rho \rangle$ are finite. 
(2) The Bose condensation does not introduce a phase factor. 
(3) The exchange term is symmetric $\langle \chi_\rho \rangle = \chi$ and real. 
Thus, these ans{\"a}tze are regarded as BCS-type weak coupling SCs. 

{\it FM}. 
Additionally, we consider the FM state. 
Here, we also introduce the local moment 
$m=\langle f_\uparrow^\dag f_\uparrow - f_\downarrow^\dag f_\downarrow \rangle$ 
as a mean-field order parameter to represent the FM long-range order. 
When this order parameter is finite, 
site-diagonal terms in the mean-field Hamiltonian have 
$ \frac{1}{4} (3J_H + J_K) m \tau_{\sigma \sigma'}^z$ for the AFM Kitaev-FM Heisenberg case 
and 
$ \frac{3}{4} J_H m \tau_{\sigma \sigma'}^z$ for the FM Kitaev-FM Heisenberg case, 
with $H_0$ modified accordingly. 
The difference between AFM Kitaev and FM Kitaev accounts for the fact 
that the FM Kitaev alone does not stabilize the FM long-range ordering 
but the AFM Kitaev coupling competes with the FM long-range ordering strongly. 
The choice of the spin axis can be taken arbitrary because of the spin rotational symmetry. 
But, with the current choice, the sixfold rotational symmetry is explicitly broken.

Except for $p$ SC$_1$, the gauge potentials $a^{x,y}=0$ while $a^z \ne 0$, 
thus the gauge symmetry is broken from $SU(2)$ to $U(1)$. 

\begin{figure*}[htbp]
\includegraphics[width=1.5\columnwidth]{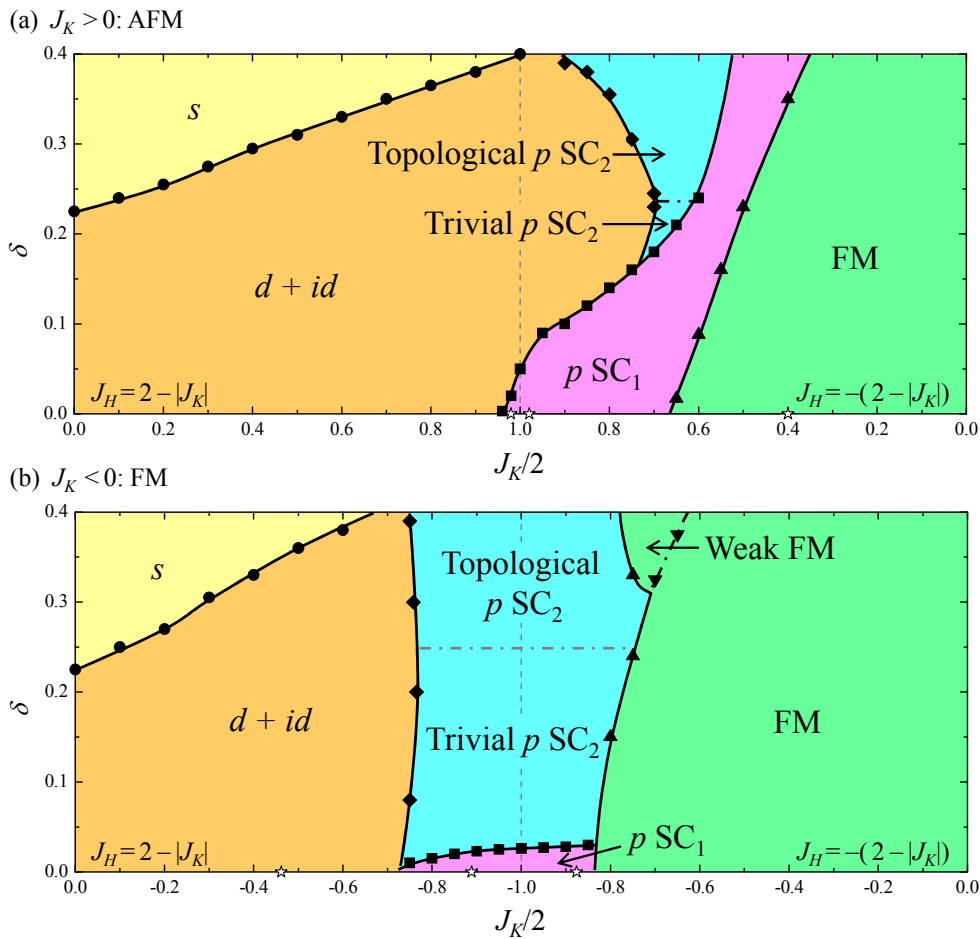}
\caption{(Color online) Mean-field phase diagrams for the doped Kitaev-Heisenberg model 
as a function of $\delta$ and $J_K$ with $J_H= \pm (2 - |J_K|)$. 
(a) AFM Kitaev ($J_K>0$) and (b) FM Kitaev ($J_K<0$). 
The phase boundary between the trivial $p$ SC$_2$ and the topological $p$ SC$_2$ in (b) (light dash-dotted line) is obtained 
with reduced SC order parameters as discussed in the main text. 
We also plot phase boundaries at $\delta=0$ obtained from the exact diagonalization 
as open stars. 
}
\label{fig:delta_jk}
\end{figure*}

\section{Results}
\label{sec:results}

\subsection{Phase diagrams}

Schematic phase diagrams for the KH model are shown in Fig.~\ref{fig:delta_jk}. 
Here, to see various phases clearly, we chose the interaction strength as $|J_K|+|J_H|=2 t$. 
In what follows, $t$ is taken as the unit of energy. 

For both the AFM Kitaev and the FM Kitaev cases, 
singlet SC states appear in the AFM Heisenberg side, $d+id$ at small $\delta$ and $s$ at large $\delta$, 
and FM states in the FM Heisenberg side. 
The difference between the AFM Kitaev and the FM Kitaev is most visible near the Kitaev limit, where 
doping-induced $p$ SC$_1$ states become unstable against the $d+id$ SC for the AFM Kitaev rather quickly 
and against the $p$ SC$_2$ for the FM Kitaev. 
The $d+id$ SC is continuously extended from the AFM Heisenberg limit, 
while the $p$ SC$_2$ for the FM Kitaev is only stable near the Kitaev limit. 
Further, the $p$ SC$_2$ for the FM Kitaev is more extended to the smaller doping regime than the $d+id$ for the AFM Kitaev. 
This difference can be understood from the different channels into which the Kitaev interaction is decoupled [Eq.~(\ref{eq:decoupling})]. 
For the AFM Kitaev, the singlet channel is weaker than the AFM Heisenberg by a factor of 3. 
On the other hand, for the FM Kitaev, the triplet channel is dominant as two components add up for one bond, 
for example $t^x$ and $t^y$ for $\gamma = z$. 
Moreover, the doping-induced kinetic energy is better gained for the $p$ SC$_1$ with the AFM Kitaev interaction 
because of the exchange term $\chi$ which is absent in the FM Kitaev interaction. 
As the AFM Kitaev interaction is decoupled into both the singlet and the triplet channels, 
the $p$ SC$_2$ could also be stabilized in the AFM Kitaev case. 
This happens when the singlet tendency is reduced by the finite FM Heisenberg interaction. 

It is noted that the phase boundary between the $p$ SC$_1$ and the FM ($d+id$ SC) for the AFM (FM) Kitaev case 
intersects the horizontal axis in the middle of the zigzag (stripy) AFM phase. 
This is expected because all states used to construct the phase diagram do not break the sublattice symmetry. 
When the zigzag and the stripy AFM states are considered, these states should also be stabilized 
near the regimes indicated by the exact diagonalization analyses. 
However, such states with longer periodicity are expected to be destabilized immediately by carrier doping 
as is the N{\'e}el AFM. 
Interestingly, the mean-field boundary between the $p$ SC$_1$ phase and the $d+id$ SC phase for the AFM Kitaev-AFM Heisenberg model 
in the limit of $\delta\rightarrow0$ agrees with the exact result on a finite cluster rather well 
[Fig.~\ref{fig:delta_jk} (a), left panel]. 
This may indicate that the uniform resonating valence bond (RVB) state at $\delta =0$ 
(singlet SC order parameters become exponentially small for both $s$ SC and $d+id$ SC states) 
is a good approximation for the N{\'e}el AFM state on a honeycomb lattice.

In Ref.~\onlinecite{Schaffer12}, 
the quantum phase transition between the Kitaev SL and the FM for the undoped FM Kitaev-FM Heisenberg model 
(or equivalently between the Kitaev SL and the stripy AFM for the undoped FM Kitaev-AFM Heisenberg model)
was studied using the SBMF approximation. 
There, the phase boundary between the Kitaev SL and the FM is shown to be located at $J_K/J_H \sim 4$, 
which is consistent with the current result [see Fig.~\ref{fig:delta_jk} (b), right panel].

In the following subsections, detailed discussions on the $p$ SC$_1$ and $p$ SC$_2$ phases and 
the relative stability between the $s$ SC and $d+id$ SC phases are presented.

\subsection{$p$ SC$_1$}

\begin{figure*}[tbp]
\includegraphics[width=2\columnwidth]{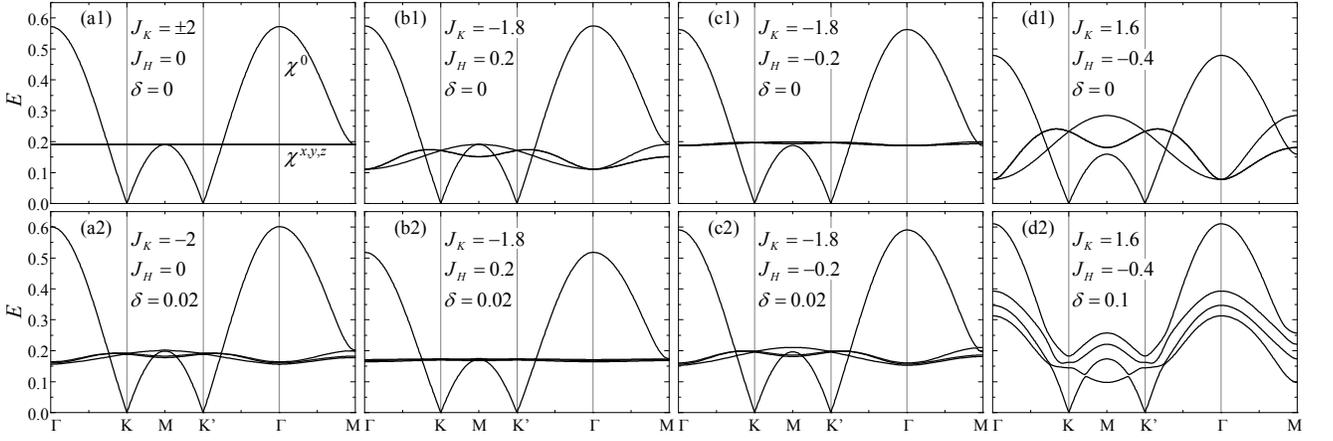}
\caption{Dispersion relations of the Majorana modes in $p$ SC$_1$ phases. 
Upper figures: undoped Kitaev-Heisenberg model; and lower figures: doped Kitaev-Heisenberg model. 
Parameter values are indicated. 
}
\label{fig:dispersion}
\end{figure*}

As discussed in Ref.~\onlinecite{You11} for the doped FM Kitaev model, 
the $p$ SC$_1$ phase is characterized by the dispersive $\chi^0$ Majorana mode and the weakly dispersive $\chi^{x,y,z}$ modes. 

Typical dispersion relations of the Majorana fermions are presented in Fig.~\ref{fig:dispersion} for various choices of parameters. 
In the undoped Kitaev limit (a1), only the gapless $\chi^0$ mode is dispersive for both the AFM and FM. 
With finite $J_H$ (b1,c1,d1), $\chi^{x,y,z}$ modes become dispersive while the $\chi^0$ mode remains gapless. 

At finite doping $\delta$, $\chi^{x,y,z}$ modes become dispersive and the $\chi^0$ mode is gapped. 
All modes are gapped by the mixing between different Majorana modes due to the finite gauge potential $a^{x,y}$. 
For the FM Kitaev interaction with $\delta=0.02$ (a2,b2,c2), 
the gap amplitude is $\sim 2 \times 10^{-6}$ and is, therefore, invisible in Fig.~\ref{fig:dispersion}. 
The finite gap in the $\chi^0$ mode results in the finite Chern number, +1 at the low doping limit. 
Softening of the $\chi^{x,y,z}$ modes is increased with the increase in $\delta$. 
However, the softening is not strong enough to close a gap for the FM Kitaev interaction 
before the $p$ SC$_1$ phase becomes unstable against the $p$ SC$_2$ phase. 
Thus, the Chern number remains $+1$.

For the AFM Kitaev interaction, 
we do see the strong softening of the $\chi^{x,y,z}$ modes (d2). 
However, gap closing needed to change the Chern number from +1 
takes place at relatively large Heisenberg interaction $|J_H/J_K| > 0.6$ and large doping $\delta > 0.1$. 
For such parameters, the current ansatz may not be a good approximation for the true ground state and/or 
the $SU(2)$ SBMF method may not be reliable.

For the FM Kitaev model, we notice that the softening of the $\chi^{1,2,3}$ modes in this work is weaker 
than that reported in Ref.~\onlinecite{You11}. 
This is supposed to originate from the level of the mean-field decoupling. 
The current decoupling is done in terms of spinons, while 
in Ref.~\onlinecite{You11} it is done in terms of Majorana fermions. 
Thus, it is possible that some order parameters, which are dropped off in the current scheme, 
are retained and have significant contributions. 
It is also noted that the $\chi^0$ mode and the $\chi^{1,2,3}$ modes are shown to overlap at the M points 
in Refs.~\onlinecite{Burnell11} and \onlinecite{Hyart12} as in the current work, while 
they do not overlap at the M points in Ref.~\onlinecite{You11}. 
Including these differences, further analyses might be necessary to fully understand the nature of the $p$ SC$_1$ phase. 

\begin{figure}[tbp]
\includegraphics[width=0.8\columnwidth]{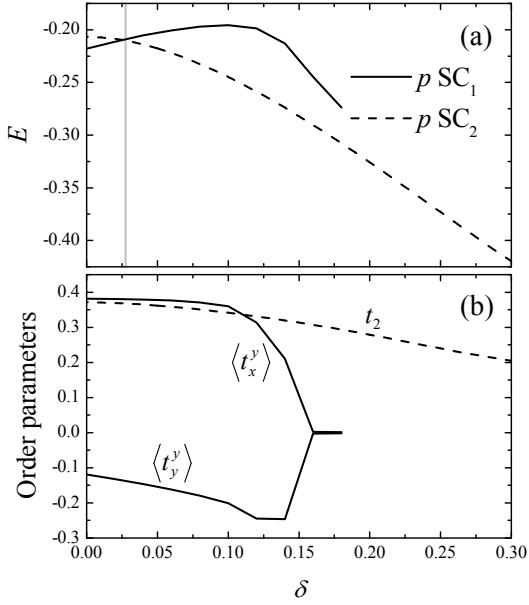}
\caption{Comparison between the $p$ SC$_1$ and the $p$ SC$_2$ for the FM Kitaev with $J_K=-2$ and $J_H=0$. 
The total energy $E$ (a) and order parameters (b) as a function of $\delta$. 
$t_2$ is defined in Eq.~(\ref{eq:t2}). 
The light vertical line in (a) indicates the boundary between the $p$ SC$_1$ phase and the $p$ SC$_2$ phase.
}
\label{fig:pSC1vspSC2}
\end{figure}

\begin{figure}[tbp]
\includegraphics[width=0.8\columnwidth]{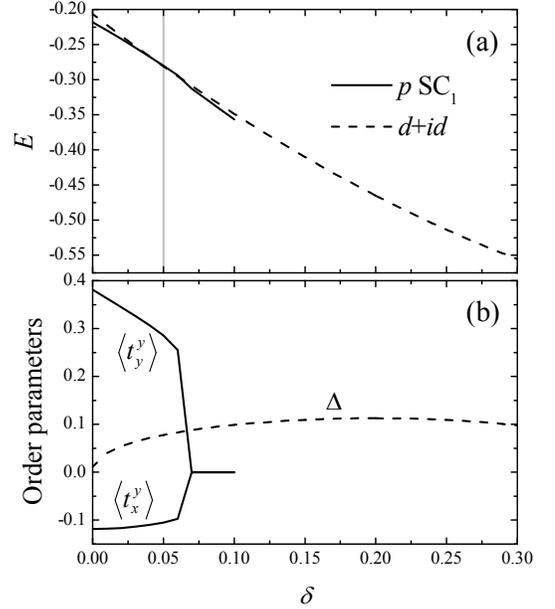}
\caption{Comparison between the $p$ SC$_1$ and the $d+id$ SC for the AFM Kitaev with $J_K=2$ and $J_H=0$. 
The total energy $E$ (a) and order parameters (b) as a function of $\delta$. 
The light vertical line in (a) indicates the boundary between the $p$ SC$_1$ phase and the $d+id$ SC phase.
}
\label{fig:pSC1vsDID}
\end{figure}

Despite the subtlety in the mean-field scheme, 
the current study provides the ``missing link'' between the previous results in 
Refs.~\onlinecite{You11} and \onlinecite{Hyart12} near the FM Kitaev limit. 
The former describes the small-doping regime correctly, while the latter describes the large-doping regime. 
Therefore, the first-order transition between the two is expected unless other phases intervene. 
In the current study, the first-order transition takes place at rather small dopings. 
The instability of the SC$_1$ phase comes from its inability to gain the kinetic energy by carrier doping 
because $\chi$ is absent in the mean-field decouplings. 
As a result, the total energy has a positive slope as shown in Fig.~\ref{fig:pSC1vspSC2}~(a). 
Similar phenomena appear to be happening in Refs.~\onlinecite{You11}; 
in Fig. 4, the order parameter $u_0$ remains constant within the SC$_1$ phase. 
On the other hand, for the AFM Kitaev case, 
the $p$ SC$_1$ phase benefits from the carrier doping like the $d$-wave SC in the $tJ$ model, and 
the total energy shows a normal behavior [see Fig.~\ref{fig:pSC1vsDID}~(a)]. 
In Figs.~\ref{fig:pSC1vspSC2}~(a) and \ref{fig:pSC1vsDID}~(a), 
one can see precursors of the unphysical behavior of the normal phase adjacent to the $p$ SC$_1$ phase; 
i.e., the sudden decrease in the total energy when the SC order parameters disappear. 
For the AFM Kitaev, this behavior starts to preempt transitions from the $p$ SC$_1$ to the $d+id$ or $p$ SC$_2$ 
by the finite FM Heisenberg interaction. 
A more reliable method such as variational Monte Carlo is necessary 
to locate the critical upper doping for the $p$ SC$_1$ phase more accurately.

\subsection{$p$ SC$_2$}

Based on the analysis on the $d$ vector,\cite{Sigrist91} 
there are three possible phases within the $p$ SC$_2$ regime: 
time-reversal symmetric (TRS) even-parity trivial phase, 
TRS odd-parity trivial phase, 
and 
TRS odd-parity topologically nontrivial or topological phase. 
In our model, all these phases could appear depending on the interaction strength and the doping concentration.

With the choice of $\theta_\gamma = 0$, 
our triplet order parameters are expressed as 
\begin{eqnarray}
&& \langle t_x^x \rangle = \langle t_y^y \rangle = \langle t_z^z \rangle = t_1, \\
&& \langle t_y^x \rangle = \langle t_z^x \rangle = \langle t_x^y \rangle = \langle t_z^y \rangle 
= \langle t_x^z \rangle = \langle t_y^z \rangle = t_2  
\end{eqnarray}
for the AFM Kitaev-FM Heisenberg model
and 
\begin{eqnarray}
\langle t_y^x \rangle = - \langle t_z^x \rangle = - \langle t_x^y \rangle = \langle t_z^y \rangle 
= \langle t_x^z \rangle = - \langle t_y^z \rangle 
= t_2 
\label{eq:t2}
\end{eqnarray}
for the AFM Kitaev-FM Heisenberg model.

\begin{figure}[tbp]
\includegraphics[width=0.8\columnwidth]{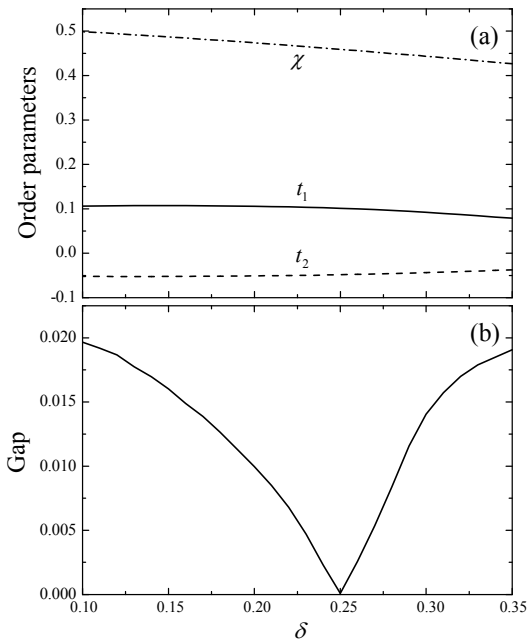}
\caption{$SU(2)$ SBMF results 
for the $p$ SC$_2$ phase in the doped AFM Kitaev-FM Heisenberg model with $J_K=1.3$ and $J_H=-0.7$. 
(a) Order parameters and (b) SC gap amplitude as a function of doping concentration $\delta$. 
}
\label{fig:GAPandOP_delta_AFK}
\end{figure}

For the AFM Kitaev case, the triplet SC order parameters are rather small as shown in Fig.~\ref{fig:GAPandOP_delta_AFK}~(a), 
and therefore the interband pairing can be neglected. 
At small dopings, there are four TR invariant $k$ points (M$_{1,2,3}$ and $\Gamma$) below the Fermi level, 
thus this SC state is in the TRS odd-parity trivial phase. 
Phase transition takes place at $\delta \sim 0.25$, 
above which only one TR invariant $k$ point ($\Gamma$) exists below the Fermi level, 
to the topologically nontrivial SC in the class DIII.\cite{Hyart12,Schnyder08} 
This transition is signaled by the gap closing with the SC order parameters remaining finite as shown in 
Fig.~\ref{fig:GAPandOP_delta_AFK}~(b). 
For the AFM Kitaev case, the choice of phases $\theta_{x,y,z}=0$ is found to correspond to 
the $d$ vector rotating around the $(1,-1,1)$ direction (see Appendix \ref{sec:dvector}). 
This corresponds to $k_x-ik_y$ pairing for spins pointing in the $(1,-1,1)$ direction and 
$k_x + ik_y$ pairing for spins pointing in the $(-1,1,-1)$ direction as in the B phase of superfluid $^3$He.  

\begin{figure}[tbp]
\includegraphics[width=0.8\columnwidth]{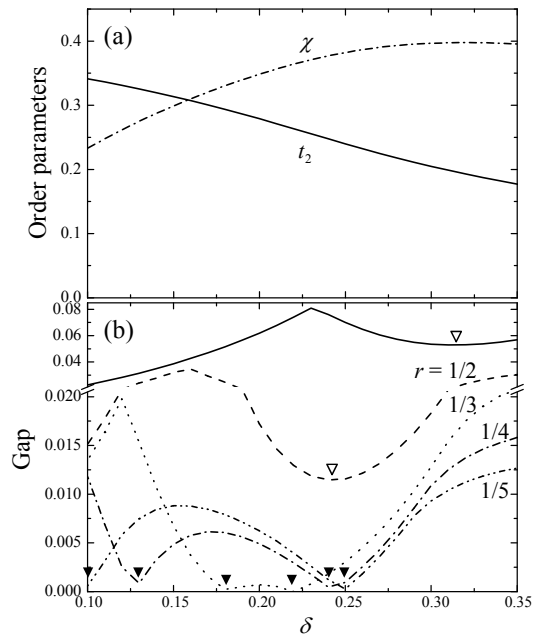}
\caption{$SU(2)$ SBMF results 
for the $p$ SC$_2$ phase in the doped FM Kitaev model with $J_K=-2$ and $J_H=0$. 
(a) Order parameters and (b) SC gap amplitude as a function of doping concentration $\delta$. 
In (b), gap amplitudes obtained by using artificially reduced SC order parameters as 
$\langle t_\rho^\gamma \rangle \Rightarrow r \langle t_\rho^\gamma \rangle$ with $r<1$
are also shown with various $r$ indicated. 
For $r \alt 0.4$, there appear two gap minima, indicating the sequence of transitions from 
the even-parity trivial phase (small $\delta$) to 
the odd-parity trivial phase (intermediate $\delta$) and 
to the odd-parity topological phase (large $\delta$). 
}
\label{fig:GAPandOP_delta_FMK}
\end{figure}

For the FM Kitaev case, the situation was found to be more complicated because the interband pairing 
has finite contributions, as the triplet SC order parameters are much larger than those in the AFM Kitaev case 
as shown in Fig.~\ref{fig:GAPandOP_delta_FMK}~(a). 
When the SC order parameters are artificially reduced as 
$\langle t_\rho^\gamma \rangle \Rightarrow r \langle t_\rho^\gamma \rangle$ with $r<1$, 
a clear transition can be seen 
between the TRS odd-parity trivial phase at $\delta <0.25$ 
and the TRS odd-parity topological phase at $\delta >0.25$ 
signaled by the gap closing [see Fig.~\ref{fig:GAPandOP_delta_FMK}~(b)]. 
As the order parameters are gradually increased, 
an additional transition shows up at small $\delta$, 
indicating the appearance of the TRS even-parity trivial phase. 
When the order parameters are fully developed, 
the TRS odd-parity trivial phase is overcome by the TRS even-parity phase, 
and the TRS even-parity phase directly transitions to the odd-parity topological phase.
Thus, as a function of temperature, the sequence of phase transition could appear within the mean-field approximation, 
although only phase transitions at zero temperature are meaningful for two-dimensional systems. 
As for the AFM Kitaev case, 
the choice of phases $\theta_{x,y,z}=0$ corresponds to 
the $d$ vector rotating around the $(-1,-1,1)$ direction in the TRS odd-parity phases.

\subsection{$s$ SC versus $d+id$ SC}

As discussed in Ref.~\onlinecite{Black07}, 
$tJ$-type models on a honeycomb lattice have some preference for the $d+id$ SC over the $s$ SC 
in the weak-coupling limit or near the critical temperature because of the interference between singlet pairing on different bonds. 
For the actual $tJ$ model excluding the double occupancy, the stabilization of the $d+id$ state was recently reported 
by using the Grassmann tensor product state approach.\cite{Gu11} 
A similar effect was observed for an electronic model with repulsive interactions.\cite{Nadkishore12} 

\begin{figure}[tbp]
\includegraphics[width=0.8\columnwidth]{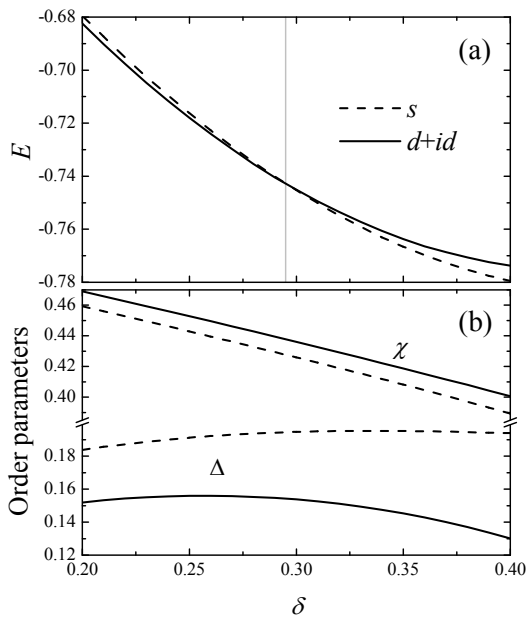}
\caption{Comparison between the $d+id$ SC and the $s$ SC for the AFM Kitaev-AFM Heisenberg model with $J_K=0.8$ and $J_H=1.2$. 
The total energy $E$ (a) and order parameters (b) as a function of $\delta$. 
The light vertical line in (a) indicates the boundary between the $d+id$ SC phase and the $s$ SC phase.
}
\label{fig:SvsDID}
\end{figure}

Within a slave-boson mean-field approach, the relative stability between $d+id$ and $s$ SC states is rather subtle.\cite{Hyart12} 
In Fig.~\ref{fig:SvsDID}, we compare the $d+id$ SC and the $s$ SC states for the doped AFM Kitaev-AFM Heisenberg model. 
As seen from the $E$-vs-$\delta$ curve, the $d+id$ SC state is stabilized at smaller $\delta$ regime, and 
the $s$ SC state is stabilized at larger $\delta$ regime. 
The $s$ SC state has the larger SC order parameter $\Delta$, 
while the $d+id$ SC state has the larger $\chi$. 
This indicates that the kinetic energy is better gained in the $d+id$ SC state, leading to its stabilization at small dopings. 

In Fig.~\ref{fig:delta_jk}, the $d+id$ SC state is shown to be stabilized near the Kitaev limit compared with the $s$ SC state. 
This is because 
the singlet pairing strength is reduced as one moves away from the AFM Heisenberg limit. 
The FM Kitaev interaction is more effective to reduce the paring strength. 
As a result, the $d+id$ SC state is extended to larger dopings. 
This consideration also explains why the $s$ SC state is extended to the lower doping regime in Ref.~\onlinecite{Hyart12}. 
There, spin-conserving exchange terms $\chi$ are not considered for mean-field order parameters. 
Thus, the kinetic-energy gain by the Heisenberg term is underestimated for the $d+id$ state.

\section{Summary and discussion}
\label{sec:summarydiscussion}

To summarize, 
we explored the possible novel phases induced by carrier doping into the KH model 
by using the $SU(2)$ SBMF method. 
Various mean-field ans{\"a}tze are motivated by the exact diagonalization results of the undoped model 
defined on a finite cluster. 
It is shown that the AFM Kitaev model and the FM Kitaev model are rather different when carriers are doped, 
although the ground state of the Kitaev model does not depend on the sign of the interaction, 
whether it is AFM or FM. 
In both cases, the $d+id$ SC state is stabilized in the AFM Heisenberg limit, 
the FM state in the FM Heisenberg limit, 
and, near the Kitaev limit, carrier doping first induces triplet superconductivity, $p$ SC$_1$. 
With the AFM Kitaev interaction, 
$p$ SC$_1$ becomes unstable against a singlet SC states with the $d+id$ symmetry, 
while with the FM Kitaev interaction it becomes unstable against another triplet SC state, $p$ SC$_2$. 
$p$ SC$_1$ state breaks the TR symmetry and has the finite Chern number; 
in the current case the Chern number +1 is rather robust. 
This state is found to be more stable with the AFM Kitaev interaction than with the FM Kitaev interaction. 
Not only for the FM Kitaev interaction, but also for the AFM Kitaev interaction the $p$ SC$_2$ state is stabilized 
when the Kitaev interaction and the Heisenberg interaction compete. 
The $p$ SC$_2$ state does not break the TR symmetry, but within this phase 
a sequence of topological phase transitions could take place. 
For the AFM Kitaev case, the intraband pairing is robust and the topological transition is between 
the TRS odd-parity trivial phase and 
the TRS odd-parity topological phase. 
On the other hand, for the FM Kitaev case, 
the interband pairing contributes when the SC order parameters are developed, 
and, depending on the magnitude of the SC order parameters, 
the topological transition could be between 
the TRS even-parity trivial phase and 
the  TRS odd-parity trivial phase, 
between the TRS odd-parity trivial phase and 
the TRS odd-parity topological phase, 
or between the TRS even-parity trivial phase and
the TRS odd-parity topological phase.

In this study, we used ans{\"a}tze which do not break the sublattice symmetry or the underlying hexagonal symmetry. 
``Zigzag'' AFM and ``stripy'' AFM phases are, therefore, 
not considered, as such complicated magnetic orderings are expected to be destabilized immediately by carrier doping. 
But it remains to be explored 
whether novel SC states are realized by carrier doping or other states outside the ans{\"a}tze are realized 
in the parameter regime where the Kitaev and the Heisenberg interactions compete.

It is an interesting and important question whether or not the present model can be realized in real materials. 
As discussed in Ref.~\onlinecite{Okamoto12}, the AFM Kitaev-AFM Heisenberg model could be realized 
in artificial TMO heterostructures, 
e.g., a bilayer of SrIrO$_3$ grown along the [111] crystallographic axis, 
when the local Coulomb interaction is large enough. 
In this case, the Heisenberg interaction is relatively large compared with the Kitaev interaction, 
and therefore the possible SC state induced by carrier doping is of the $d+id$.

For (topological) quantum computations, triplet SC states, $p$ SC$_1$ or $p$ SC$_2$ in the nontrivial phase, 
are desired. 
To realize the topological $p$ SC$_2$ state, 
one should include 
the FM Kitaev interaction as the dominant interaction  
or the AFM Kitaev interaction with finite FM Heisenberg interaction to suppress the tendency towards the singlet formation. 
$A_2$IrO$_3$ with $A$=Li or Na was originally suggested as a candidate for realizing the FM Kitaev interaction. 
But, later it was experimentally shown to have zigzag AFM ordering, 
indicating the importance of the longer-range interaction or 
the Kitaev interaction is actually AFM with the finite FM Heisenberg interaction. 
If the latter situation is realized, carrier doping may induce triplet SCs. 
Yet, even in this case, the carrier hopping term does not conserve the isospin. 
Therefore, the stability of the triplet SC states depends on the strength of the isospin-nonconserving hopping.

\acknowledgements
We thank G. Khaliullin and R. Thomale for their fruitful discussions and comments. 
This research was supported by 
the U.S. Department of Energy, Basic Energy Sciences, Materials Sciences and Engineering Division.

\appendix

\section{Stability matrix for the $p$ SC$_2$ phase}
\label{app:stabilitymatrix}

The symmetry of the superconducting order parameters at the critical temperature $T_c$ 
can be deduced by analyzing the stability matrices\cite{Hyart12,Black07} 
which are derived from the linearized gap equations. 
For the triplet superconductivity $p$ SC$_2$, 
the stability matrices consist of three independent matrices corresponding to 
$\langle t^x_\rho \rangle, \langle t^y_\rho \rangle$ and $\langle t^z_\rho \rangle$. 
For $\langle t^x_\rho \rangle$, the stability matrix $M_x$ is given by 
\begin{eqnarray}
M_x = 
\left[ 
\begin{array}{ccc}
(J_K -J_H) B & -J_H C & -J_H C \\
(J_K -J_H) C & -J_H B & -J_H C \\
(J_K -J_H) C & -J_H C & -J_H B 
\end{array}
\right] 
\label{eq:Mx1}
\end{eqnarray}
for the AFM Kitaev-FM Heisenberg model 
and 
\begin{eqnarray}
M_x = 
\left[ 
\begin{array}{ccc}
-J_H B & -(J_K + J_H) C & -(J_K + J_H) C \\
-J_H C & -(J_K + J_H) B & -(J_K + J_H) C \\
-J_H C & -(J_K + J_H) C & -(J_K + J_H) B 
\end{array}
\right] 
\label{eq:Mx2}
\end{eqnarray}
for the FM Kitaev-FM Heisenberg model. 
Here, $B=A_{\rho = \rho'}$ and $C=A_{\rho \ne \rho'}$, with the matrix $\hat A$ given by 
\begin{eqnarray}
A_{\rho \rho'} \!\!&=&\!\! \frac{1}{2} \sum_{\vec k} \biggl[ \biggl( 
\frac{\tanh (\varepsilon_+/2 k_B T_c)}{2 \varepsilon_+}
+
\frac{\tanh (\varepsilon_-/2 k_B T_c)}{2 \varepsilon_-}
\biggr) \nonumber \\
&& \times 
\sin (\vec k \cdot \vec r_\rho - \theta) \sin (\vec k \cdot \vec r_{\rho'} - \theta)
\nonumber \\
&&+
\frac{\sinh (\mu /k_B T_c) \cos (\vec k \cdot \vec r_\rho - \theta) \cos (\vec k \cdot \vec r_{\rho'} - \theta)}
{2 \mu \cosh (\varepsilon_+/2k_B T_c) \cosh (\varepsilon_-/2k_B T_c)}
\biggr]. \nonumber \\
\end{eqnarray}
Considering a symmetric state with $\langle \chi \rangle$ being independent of the bond specie, 
$\varepsilon_\pm$ is given by 
$\varepsilon_\pm = \pm |\varepsilon (\vec k)|-\mu$ and $\theta = \arg [\varepsilon (\vec k)]$ 
with 
$\varepsilon (\vec k) = - t_{eff} \sum_\rho e^{i \vec k \cdot \vec r_\rho}$. 
Here, $t_{eff} = \frac{1}{2} \delta t + \frac{1}{8} 
\{ J_K \Theta (J_K) + 3 J_H \Theta (J_H) \}  \langle \chi^* \rangle$ with 
$\Theta$ being the Heaviside function. 
The leading pairing instability is determined by 
the eigenvector with the largest eigenvalue of Eq.~(\ref{eq:Mx1}) or (\ref{eq:Mx2}). 
As $C<0$, such an eigenvector is expressed as 
$\mbox{\boldmath $d$}_x = \langle t^x_x, t^x_y, t^x_z \rangle = (\sqrt{1-2 \eta^2},-\eta^2,-\eta^2)$ for the AFM Kitaev and 
$\mbox{\boldmath $d$}_x = (0, 1/\sqrt{2},-1/\sqrt{2})$ for the FM Kitaev. 
The stability matrices $M_{y,z}$ and the eigenvectors for $M_{y,z}$ with the largest eigenvalue, 
say $\mbox{\boldmath $d$}_y$ and $\mbox{\boldmath $d$}_z$, 
can be obtained from $M_x$ and $\langle t^x_x, t^x_y, t^x_z \rangle$, respectively, 
by cyclically exchanging components. 
Any linear combinations of $\mbox{\boldmath $d$}_{x,y,z}$ give the same critical temperature. 
But, the stable pairing amplitude at low temperatures must be determined by solving the non-linear gap equations.

\section{$d$ vector analysis for the $p$ SC$_2$ phase} 
\label{sec:dvector}

Here, we consider both intraband $d$ vectors ($\mbox{\boldmath $d$}_{11}$) and 
interband $d$ vectors ($\mbox{\boldmath $d$}_{12}$) (Refs.~\onlinecite{Sigrist91,Black07}) 
for our doped KH models 
by expanding the exponents in the anomalous terms [Eqs.~(\ref{eq:DeltaAFKitaev}) and (\ref{eq:DeltaFMHeiisenberg})] 
in the mean-field Hamiltonian around $\vec k =0$. 
For the AFM Kitaev-FM Heisenberg model, the intraband pairing is found to be dominant and the $d$ vector is given by 
\begin{eqnarray}
\mbox{\boldmath $d$}_{11} 
\!\!&=&\!\! i D \biggl( 
-\frac{\sqrt{3}}{2} k_x - \frac{1}{2} k_y,
-\frac{\sqrt{3}}{2} k_x + \frac{1}{2} k_y, 
 k_y \biggr) , \label{eq:dintra} 
\end{eqnarray}
where 
$D = \frac{1}{8} \{ (J_K - J_H) t_{1} + J_H  t_{2} \}$ 
with $t_{1 (2)} = \langle t_\rho^\gamma \rangle$ for $\gamma = (\ne) \rho$. 
For the FM Kitaev-FM Heisenberg case, 
using the same procedure for Eqs.~(\ref{eq:DeltaFMKitaev}) and (\ref{eq:DeltaFMHeiisenberg}), 
we obtain 
\begin{eqnarray}
\mbox{\boldmath $d$}_{11} 
\!\!&=&\!\! i D \biggl( 
\frac{1}{2} k_x - \frac{\sqrt{3}}{2} k_y,
-\frac{3}{2} k_x - \frac{\sqrt{3}}{2} k_y, 
 - k_x \biggr) , \label{eq:dintra} \\
\mbox{\boldmath $d$}_{12} 
\!\!&=&\!\! \frac{1}{2} D \biggl( 
-\frac{\sqrt{3}}{4} k_x^2 + \frac{1}{2} k_x k_y + \frac{\sqrt{3}}{4} k_y^2, \nonumber \\
&& \hspace{2em} \frac{\sqrt{3}}{4} k_x^2 + \frac{1}{2} k_x k_y - \frac{\sqrt{3}}{4} k_y^2, 
- k_x k_y \biggr) , 
\end{eqnarray}
where 
$D=\frac{\sqrt{3}}{8} (J_K+J_H) t_2$ with 
$t_2 = \langle t_y^x \rangle = -\langle t_z^x \rangle = - \langle t_x^y \rangle = \langle t_z^y \rangle 
= \langle t_x^z \rangle = - \langle t_y^z \rangle$.

When the intraband pairing is dominant, the $p$ SC$_2$ is in the TRS odd-parity phase. 
The choice of $\theta_{x,y,z}=0$ above 
describes the $d$ vector rotating around the $(1,-1,1) [(-1,-1,1)]$ direction for the AFM (FM) Kitaev case. 
This corresponds to $k_x-ik_y$ pairing for spins pointing in the $(1,-1,1) [(-1,-1,1)]$ direction and 
$k_x + ik_y$ pairing for spins pointing in the $(-1,1,-1) [(1,1,-1)]$ direction 
as in the B phase of superfluid $^3$He.  
For the FM Kitaev case, the contribution from the interband pairing becomes large when the SC order parameters are developed, 
resulting in the TRS even-parity phase in the small-doping regime.

\end{document}